\documentclass{epl}
\usepackage{graphicx}

\newcommand{\be}{\begin{equation}}
\newcommand{\ee}{\end{equation}}
\newcommand{\ba}{\begin{eqnarray}}
\newcommand{\ea}{\end{eqnarray}}

\newcommand{\half}{\frac{1}{2}}

\title{A practical density functional for polydisperse polymers}
\shorttitle{A density functional for polymers}
\author{I. Pagonabarraga\inst{1,2} \and M. E. Cates\inst{1}}
\institute{
\inst{1} Department of Physics and Astronomy, University of Edinburgh, Kings Buildings, Mayfield Road, Edinburgh EH9 3JZ, Scotland\\
\inst{2} Departament de F\'{\i}sica Fonamental, Universitat de Barcelona, \\
Av. Diagonal 647, 08028-Barcelona, Spain}

\begin{document}
\maketitle
\begin{abstract}
The Flory Huggins equation of state for monodisperse polymers can be turned into a density functional by adding a square gradient term, with a coefficient fixed by appeal to RPA (random phase approximation). We present instead a model nonlocal functional in which each polymer is replaced by a deterministic, penetrable particle of known shape. This reproduces the RPA and square gradient theories in the small deviation and/or weak gradient limits, and can readily be extended to polydisperse chains. The utility of the new functional is shown for the case of a polydisperse polymer solution at coexistence in a poor solvent. 
\end{abstract}
\section{Introduction}
Recent work on polydisperse thermodynamics has shown the value of using, as thermodynamic coordinates, whatever linear combinations of densities actually occur in the excess free energy of the system. These linear combinations are `generalized moments' (moments for short) of the complete set of densities, which are in principle infinite in number; however, in most models, particularly in polymeric systems, they form a much reduced set. For example, the Flory Huggins model of length-polydisperse homopolymers and solvent on a unit lattice has (setting $kT = 1$) $F_{ex} = (1-\phi)\ln(1-\phi)+\chi \phi(1-\phi)$, where $\phi = \int dN N\rho(N)$ is the monomer density, or first moment of the distribution $\rho(N)$ that specifies the density of chains of length $N$. Thus there is only one moment in $F_{ex}$. 

It was shown \cite{warren,sollich,long} how to construct a `projected' free energy, expressed as a function of the required moments, whose predictions for cloud curves, shadow curves, and critical points coincide with those of the full model. It was also shown \cite{sollich,long} how, by systematically adding additional moments, the full phase behaviour (describing states of coexistence midway along a tie-ine as well as at its endpoints) could be recovered to any desired accuracy. 

An important extension of this approach is to address interface thermodynamic behaviour, for example to predict the interfacial tensions between coexisting phases. (In doing this, we can suppose that the {\em composition} of such phases was calculated already, {\it e.g.}, by the methods just outlined.) This requires a model, not merely of the equation of state, but of the free energy functional. In some cases (such as polydisperse hard spheres) accepted functionals include ones where the excess free energy depend only on a few nonlocal `weighted densities'. The minimization equations then close in the subspace of these weighted densities \cite{pagona}, simplifying practical calculation. So far, there is no such treatment of polydisperse polymers, despite the simplicity of the excess free energy. This is partly because polymers are extended objects: the main difficulty lies in generalizing the {\em ideal} (entropic) term to inhomogeneous states.

\section{RPA and Square Gradient Models}
For concentrated monodisperse (homo)polymers, there are two simple methods for generating density functionals. The first is to expand in the random phase approximation (RPA) for {\em small deviations} about a uniform reference state. In Fourier space this gives to quadratic order
\begin{equation}
\Delta F = \half\sum_{q\neq 0} |\phi(q)|^2 \left(S_0(q)^{-1} +  F_{ex}''(\phi_0)\right) 
\label{eqn:RPA}
\end{equation}
\begin{equation}
S_0(q) = \phi_0 s_0(q) = \rho_0 N^2g_D(q^2R_g^2)
\label{eqn:structure}
\end{equation}
Here vector notation for the wavevector is suppressed and $F_{ex}''$ is evaluated in a uniform state, of monomer density $\phi_0 = \rho_0N$. The unperturbed scattering function $S_0(q)$ describes an ideal gas of Gaussian chains (each with structure factor $s_0(q)$) at number density $\rho$. In eq.\ref{eqn:structure}, $g_D(x) = 2(\exp(-x)-1+x)/x^2$ is the Debye function and $R_g$ the chain gyration radius (on a unit lattice, $(N/6)^{1/2}$) \cite{footfactor}. Within RPA, one can continue to higher order, but with increasingly complicated results \cite{RPA}. A disadvantage of eq.\ref{eqn:RPA} is that it does not allow study of (say) the interface between phases at coexistence, whose composition difference is finite.

The second (square gradient) method \cite{jodeg} allows for such finite deviations but assumes instead {\em weak gradients}, asserting that (again suppressing vector notation)
\begin{equation}
\Delta F = \int \left[F(\phi(r)) + K(\phi) |\nabla\phi|^2\right]\,dr
\label{eqn:squaregrad}
\end{equation}
where $F(\phi) = F_{id}+F_{ex}(\phi)$ and $F_{id}=(\phi/N)[\ln(\phi/N)-1]$ is the ideal contribution to the Helmholtz free energy of a homogenous phase. The coefficient $K(\phi)$ is {\em a priori} unknown, but can be identified by comparing with eq.\ref{eqn:RPA} at order $q^2$. This gives (on a unit lattice) $K^{-1} = 36\phi$.  This choice for $K$ ensures that RPA is recovered when $\phi$ variations are small as well as slowly varying. For strong segregation (when neither applies) some authors (but not others \cite{critique}) propose matching instead the coefficient of $|\nabla\phi|^2$ against the high-$q$ limit of $S_0(q)^{-1}$ which is also quadratic \cite{broseta}; this gives $K^{-1} = 24\phi$. 

\section{An Alternative Functional}
The implementation of either of these methods for polydisperse chains is less simple than it looks (see \cite{warrenpccp}, and below). We offer instead a density functional for dense polymer systems which reduces to the RPA and square gradient functionals in the appropriate limits, but whose extension to polydispersity is both transparent and simple to implement. In effect, we replace each chain by a nondeformable `effective particle' with extremely non-pairwise-additive interactions. (This last feature contrasts with, and complements, a related recent approach to {\em dilute} polymers \cite{hansen}.) Our functional interpolates, in an illuminating way, the physics of RPA and of the square gradient limits.

For a single polymeric species in a monomeric solvent, we propose the model functional
\begin{equation}
\Delta F = \int \left\{\rho(r)[\ln \rho(r)-1]+ F_{ex}(\phi(r))\right\} \,dr
\label{eqn:monofunctional}
\end{equation}
\begin{equation}
\rho(q) = (N s_0(q))^{-1/2}\,\phi(q)
\label{eqn:convol}
\end{equation}
Expanding this functional in small deviations about a uniform reference state recovers immediately eq.\ref{eqn:RPA}, whereas expanding it in gradients yields eq.\ref{eqn:squaregrad}. The latter expansion is achieved by writing $\rho(r) = (1/N)(1-(N/36)\nabla^2)\phi(r)$ (this is, in real space, the low $q$ expansion of eq.$\ref{eqn:convol}$), substituting into eq.\ref{eqn:monofunctional} and integrating by parts. The result $K^{-1} = 36\phi$ is recovered without further assumption.
These facts are sufficient to ensure that our functional encompasses, for monodisperse chains, the standard (leading order) RPA and square-gradient functionals. 

\section{Interpretation} 

Eq.\ref{eqn:monofunctional} has a simple physical interpretation, as follows.  
It describes a fluid of identical, interpenetrable, spherical particles of fixed structure.
The density of the {\em centres} of these particles is $\rho(r)$ and the {\em ideal} part of the free energy functional is simply $\rho(r)[\ln\rho(r)-1]$ (as applies for any simple fluid). Viewing each particle as a nondeformable cloud of monomers, it follows that the monomer density $\phi$ is constructed from $\rho$ by a deterministic, invertible real-space convolution, or Fourier product (eq.\ref{eqn:convol}): $\phi(q) = w(q)\rho(q)$. The kernel $w$ is, remarkably, not the Debye function for a Gaussian chain, but proportional to its square root: $w(q) = N g_D^{1/2}$. This choice ensures that the monomer-monomer correlator (in Fourier space, $S_0(q)$) in an ideal gas of our penetrable spheres is identical to that of an ideal gas of Gaussian chains. The kernel $w$ itself is {\em not} the monomer-monomer correlator of an effective particle, but (in real space) the conditional probability of finding a monomer at $r$ given that the {\em centre} of our effective particle is at the origin \cite{footform}. Each effective particle is a spherically pre-averaged polymer in the sense that it has the same monomer-monomer correlator; but the density profile about its centre is not that about any point on a real chain (nor the centre of mass). Indeed, since a polymer is stochastic, one cannot have both things at once; we give priority to $S_0(q)$. Finally (as in both the RPA and the square gradient models) $F_{ex}$ is taken as a local functional of the monomer density $\phi$. This reflects the fact that this term arises from enthalpic interactions unrelated to chain connectivity, whose scale is local (monomeric) \cite{jodeg}.

Eq.\ref{eqn:monofunctional} illustrates the central issue in constructing 
density functionals for extended objects. Within Flory Huggins, the entropy in 
any homogeneous state is that of an ideal gas of chains. But a chain is not 
pointlike, and sees variations in density -- so, which density should we use? 
The `effective particle' description gives a definite (though perhaps not unique)
 answer. If eq.\ref{eqn:monofunctional} is written down only in terms of
 $\phi$, the entropy term is nonlocal \cite{RPA}. If written only in terms 
of $\rho$, then the enthalpic term is nonlocal instead. This `uncertainty 
principle' will hold for any theory of extended objects with local interactions. 
Both $\rho$ and $\phi$ are needed, and our approximation lies in adopting a 
fixed relation between them, instead of the actual one which for polymers is
 (a) stochastic and (b) subject to perturbation by strong density variations. 

Our effective particles cannot adapt to their surroundings by deformation, only by displacement; this limits the formal range of validity for eq.\ref{eqn:monofunctional} mainly to cases where either RPA or the square gradient theory could have been validly used instead. On the other hand, by adopting a square gradient form also in strong segregation \cite{broseta}, the nondeformability of our particles could be relaxed, which {\em might} better approximate the behaviour of real polymers (to the same extent that strong segregation would do). In what follows we consider only weak and intermediate ($R_g\nabla\simeq 1$) segregation where eq.\ref{eqn:monofunctional} is at least competitive.

\section{Polydispersity}
The present approach comes into its own in the presence of polydispersity. The underlying `effective particle' interpretation makes the required extension both obvious and tractable. To model length-polydisperse polymers at number density $\rho(N)$ (where $N$ is now treated as a continuous variable) and with an excess free energy that depends (as before) only on the local monomer density $\phi(r)$, we must choose 
\begin{equation}
\Delta F = \int dr \left\{\int\rho(N,r)[\ln \rho(N,r)-1]\,dN + F_{ex}(\phi(r))\right\}
\label{eqn:polyfunctional}
\end{equation}
\begin{equation}
\phi(r) \equiv \int \phi(N,r)\,dN
\label{eqn:chainsum}
\end{equation}
\begin{equation}
\phi(N,r) = \int w(N,r-r')\rho(N,r')\,dr'
\label{eqn:convol2}
\end{equation}
where, in Fourier space, $w(N,q) = N g_D(q^2N/6)^{1/2}$, as before. The convolution between $\phi(N,r)$ (the monomer density at $r$ from objects with $N$ monomers) and $\rho(N,r)$ (the density of centres of these objects) remains invertible; thus eq.\ref{eqn:convol2} directly generalizes eq.\ref{eqn:convol}.

Eq.\ref{eqn:polyfunctional} allows practical numerics, because it is in the class where the excess free energy depends on only a few `weighted densities' or nonlocal linear combinations of the particle densities $\rho(N,r)$. In such cases, it is possible (once the bulk chemical potentials are known -- equivalent to specifying the bulk composition $\rho_0(N)$ of any one of the phases present) to exactly obtain autonomous (closed) equations among these weighted densities \cite{pagona}. The size distribution at any point in the density profile can be calculated afterwards. 

In the present case there is only one weighted density ($\phi$) with the autonomous equation:
\begin{equation}
\phi(r) = \int \int w(N,r-r')\left[\rho_0(N)\exp\left(-\int w(N,r''-r')\Delta F_{ex}'(\phi(r''))\,dr''\right)\right]\,dN\,dr'
\label{eqn:autonomous}
\end{equation}
where $\Delta F_{ex}'(\phi) = F'_{ex}(\phi)-F'_{ex}(\phi_0)$ in which the 
prime denotes a derivative; $\phi_0$ is the bulk monomer density in 
the chosen reference phase. Note that $\Delta F_{ex}'(\phi)$ is a function 
only of $\phi$, hence the closure at this level; however the integral over
$N$ means that eq.\ref{eqn:autonomous} is not (to our knowledge) the Euler-Lagrange equation from any functional of $\phi(r)$ alone.

\section{Example}

\begin{figure}
\begin{center}
\includegraphics[height=8cm]{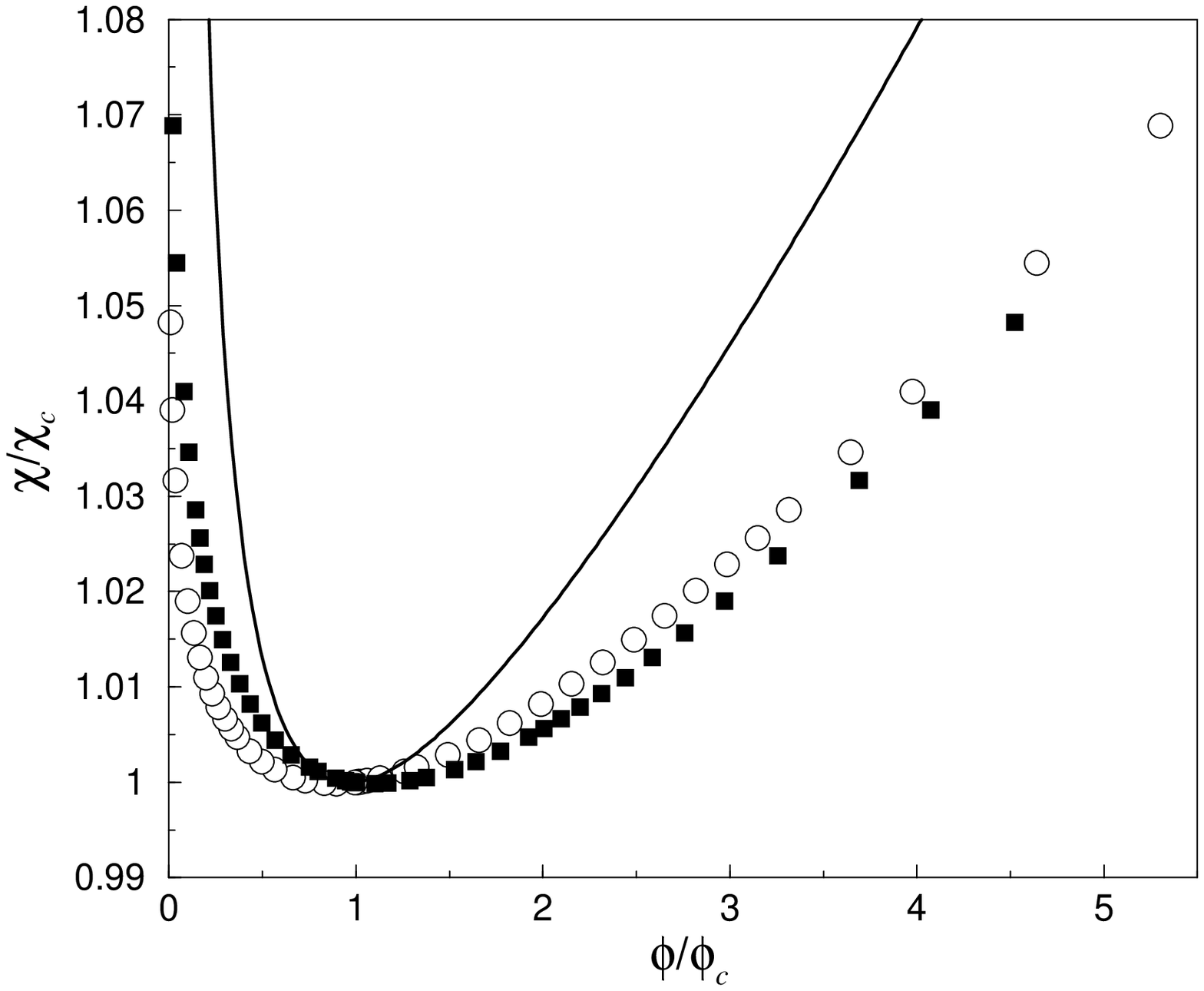}
\includegraphics[height=8cm]{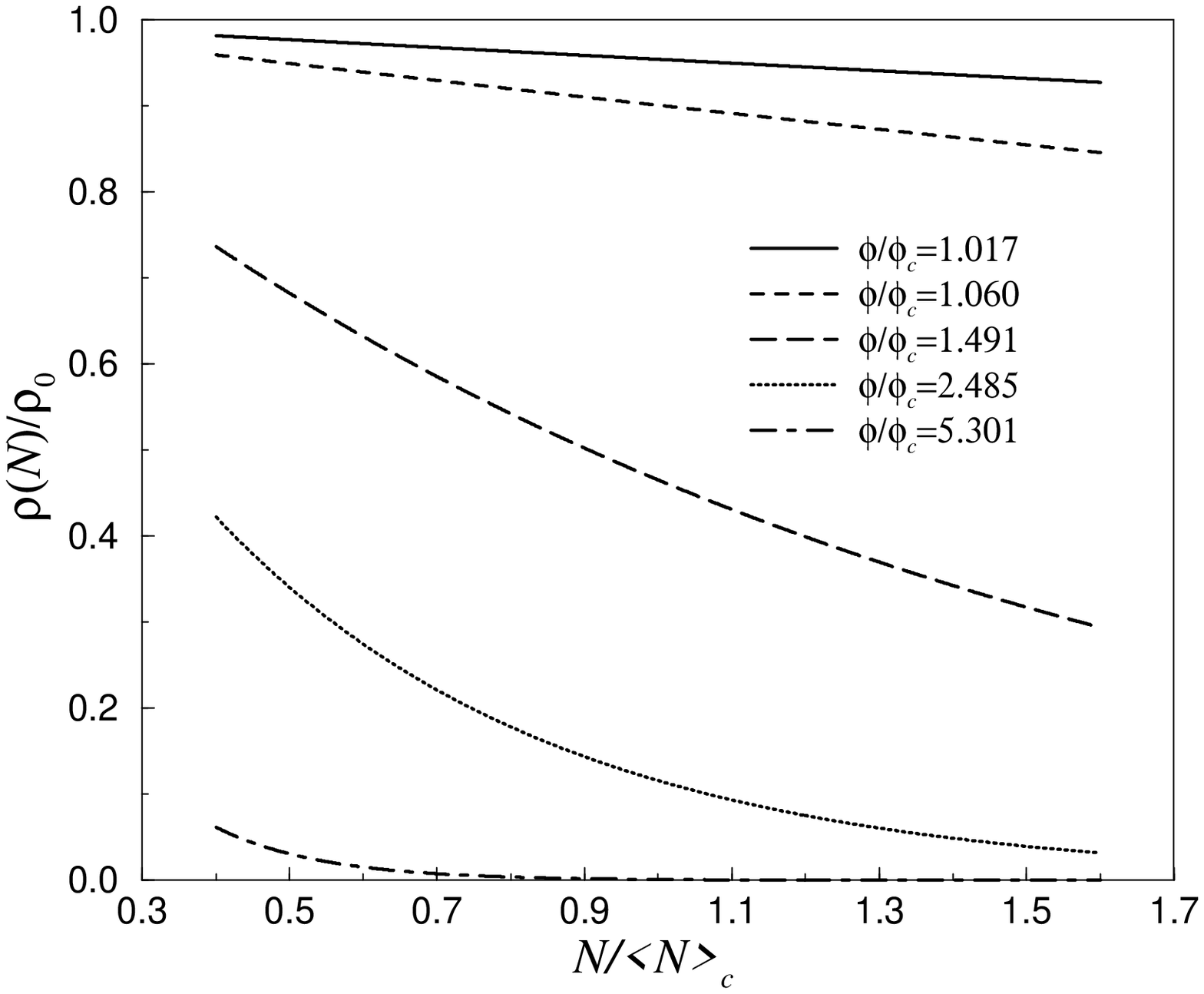}
\end{center}
\caption{{\it (a)} Cloud (circles) and shadow (filled squares) curves for a uniform chain length
distribution. The continuous curve is the spinodal. {\it (b)} Composition 
in the shadow phase for various points on the cloud curve (identified by $\phi$ in the cloud phase, whose dependence on $\chi$ is shown in {\it (a)}).}
\label{fig:eq}
\end{figure}

An example follows of a practical application of the method, generating interfacial profiles and properties for coexisting phases of polydisperse homopolymer in a solvent 
. We have considered, for simplicity, the case where 
the parent phase has a uniform length distribution (with mean length $\langle N\rangle_0 = 1000$ 
and standard deviation 346 in the parent). We study coexistence between a cloud phase and its shadow, solving eq.\ref{eqn:autonomous}; the cloud phase is then the parent or reference $\rho_0(N)$ 
and we find the interface between it and the (unique) shadow phase at a given $\chi$ parameter. (By using an `effective parent' the same method extends to any point along a tie line; see \cite{long}.) In fig.\ref{fig:eq}a, we show the relevant coexistence curves, which display a dense polymer solution (cloud) coexisting with a dilute one (shadow). The length distribution in 
the shadow differs from the (flat) distribution in the cloud, as shown in 
fig.\ref{fig:eq}b; for our flat parent the size distribution in the shadow is exponential. The shorter chains are preferentially segregated into the shadow phase, which, in addition, has fewer chains overall.

\begin{figure}
\begin{center}
\includegraphics[height=8cm]{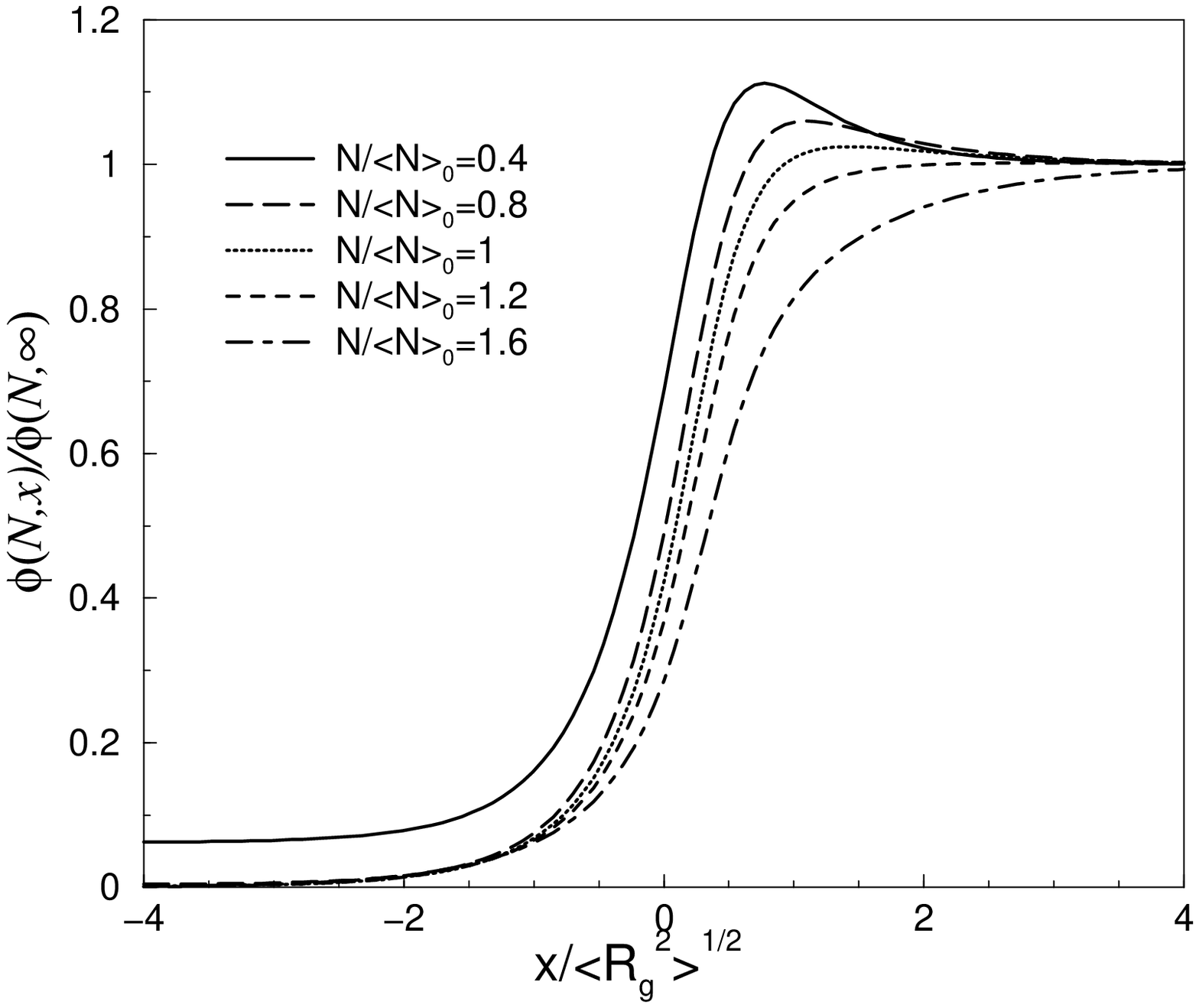}
\includegraphics[height=8cm]{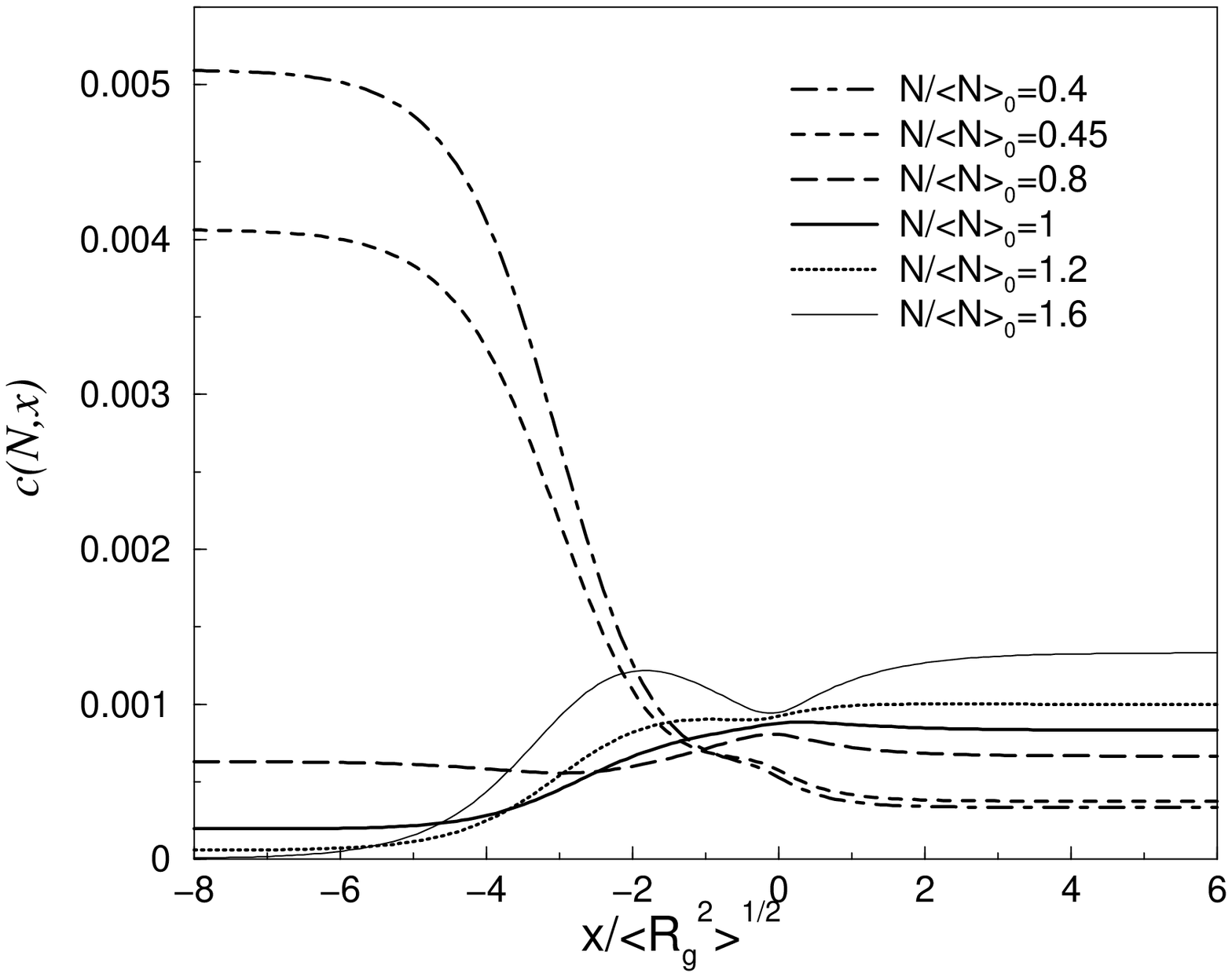}
\end{center}
\caption{Monomeric density profiles for various of species for a flat parent at its cloud point, with $\langle \phi \rangle_0=0.16$ and $\chi/\chi_c = 1.07$. {\it (a) }
Monomer concentration profiles, each normalized by its value in the cloud phase. 
{\it (b)} Relative concentration profiles (for the same species).}
\label{fig:profs}
\end{figure}

We have solved numerically eq.\ref{eqn:autonomous}, which gives us access 
to both structural and thermodynamic quantities for the interface between coexisting phases. For numerical convenience, the Debye function was replaced by $g(x) =
[1+x^2/2]^{-1}$ as is standard practice \cite{doi}. The first moment density (or total monomer concentration) 
exhibits smooth, monotonic profiles. From these, we can define 
an interfacial width, which shows a smooth decrease when 
increasing the interaction strength, $\chi$. However, at the same time we find a nonmonotonic distribution of monomer concentrations for individual species when we move from the vicinity of the critical point. In fig.\ref{fig:profs}a, we show the monomer concentration profiles contributed by various different chain lengths, centered at the position where $\phi(r)$ takes its mean value. The shorter polymers show a maximum in the monomer concentration at the interface: they act rather like a surfactant. For the same selection of species, we show in fig.\ref{fig:profs}b profiles
for the relative amounts of each species, $c(N,x) = \phi(N,x)/\phi(x)$. The strong depletion of the long chains in the dilute phase induces a nonmonotonic, even oscillatory, behaviour in this quantity. Closer to the critical point, 
the profiles for the monomer and relative concentrations change monotonously 
from the dense to the dilute phase. 

\begin{figure}
\includegraphics[height=8cm]{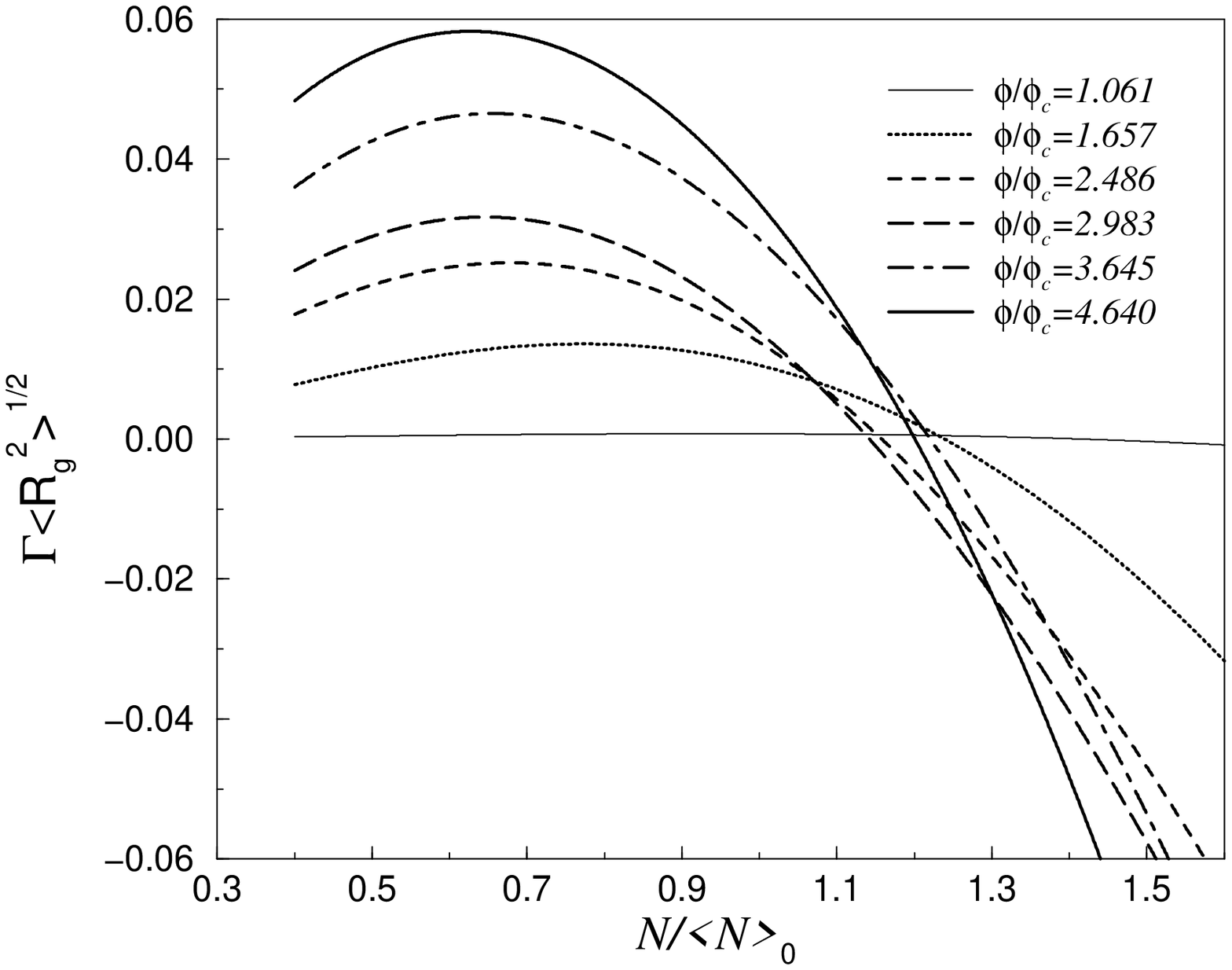}
\includegraphics[height=8cm]{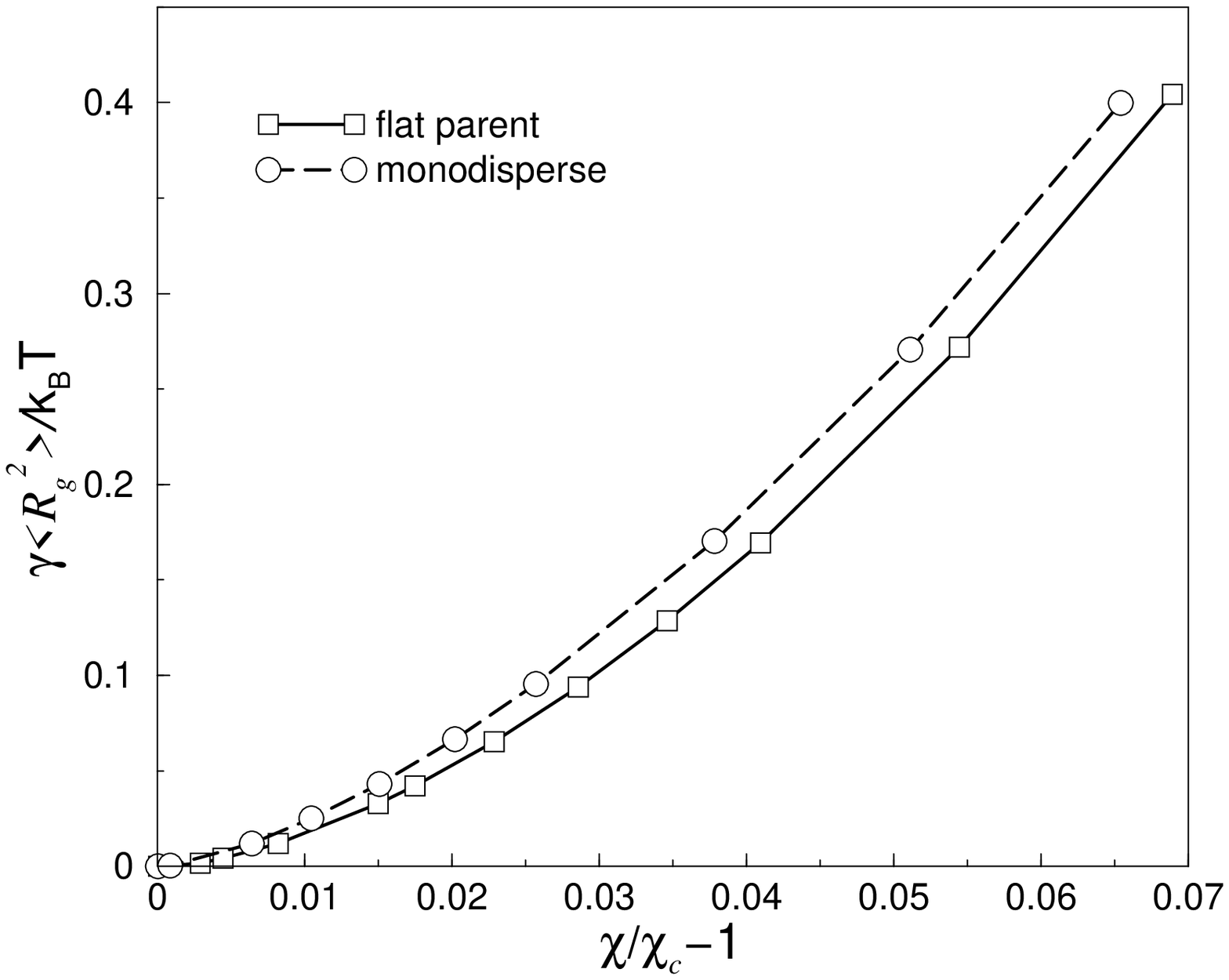}
\caption{{\it (a)} Monomeric adsorption $\Gamma$, as a function of chain length, for the same parent as in the previous figures, at various points on the cloud curve. {\it (b)} Surface tension for the flat parent and an equivalent  monodisperse solution, {\it i.e.} with the same $<N>_0$.}
\label{fig:ads}
\end{figure}

In fig.\ref{fig:ads}a we show the 
monomeric adsorption as a function of chain length for a fixed parent at various
 $\phi$ values. The adsorbed amounts are defined with respect to the Gibbs dividing
surface of the overall monomer concentration. The smaller chains are adsorbed 
preferentially (consistent with fig.\ref{fig:profs}a), and the species that has
the maximum adsorption corresponds to a smaller chain for stronger demixing. 
The latter can be attributed to an increased skewness of the shape distribution 
in the dilute phase \cite{pago2}. Finally, in  fig.\ref{fig:ads}b we compare the surface 
tension of the polydisperse fluid with an equivalent monodisperse one. It shows 
the surfactant effect of the short chains -- the surface tension of the 
polydisperse mixture increases more slowly than its monodisperse counterpart -- though the effect is surprisingly weak. A more detail analysis of these and the other results will be pursued elsewhere \cite{pago2}.

\section{RPA and Square Gradient Limits}
Having shown its utility by this example, we return to the general properties of our functional. Expanding in small deviations (RPA) now allows two separate descriptions, according to whether the densities of individual species are eliminated or retained. The simplest RPA is to keep eq.\ref{eqn:RPA} but to replace eq.\ref{eqn:structure} with the ideal structure factor of the polydisperse noninteracting system (see, {\it e.g}. \cite{huang}). This is surely correct, but allows no calculation of partial structure factors for differently labelled species and cannot be extended to dynamics \cite{warrenpccp}, or even interfaces.

The full RPA limit of our functional is instead
\begin{equation}
\Delta F = \half\sum_{q\neq 0} \int |\phi(N,q)|^2 \left(S_0(N,q)^{-1} + F_{ex}''(\phi_0)\right)\,dN 
\label{eqn:RPAfull}
\end{equation}
This coincides with the second of two distinct polydisperse RPAs suggested by Warren \cite{warrenpccp}; for our effective particles (at least) this is rigorous, and his first suggestion, which treats $\rho(N,r)$ and $\phi(N,r)$ as though independent, incorrect.

Turning to the square gradient theory, we obtain from our functional (after expansion and integration by parts)
\begin{equation}
\Delta F = \int dr \int
\left\{
\left(\frac{\phi(N,r)}{N}\right)
\left[\ln\left(\frac{\phi(N,r)}{N}\right)-1\right] + \frac{1}{36\phi(N,r)}(\nabla\phi(N,r))^2 
\right\}\,dN 
+ F_{ex}(\phi(r))
\label{eqn:fullsquare}
\end{equation}
as used, for example, in \cite{clarke}. This is certainly the proper extension of eq.\ref{eqn:squaregrad}, although the fact that the resulting equations can be closed in $\phi(r)$ has, we think, been overlooked previously. In fact, at square gradient level eq.\ref{eqn:autonomous} is converted from an integral equation into a second order differential equation for $\phi(r)$. However, this is so nonlinear that it offers little numerical advantage over eq.\ref{eqn:autonomous} and we found it just as easy to use the full functional. Analytically, the expansion in gradients does allow some limiting trends to be extracted \cite{pago2}; a square-gradient truncation might also be used to mimic chain deformability in the strong segregation limit \cite{broseta}.

\section{Extensions}
No assumption about $F_{ex}$ was made other that it was a local function of the overall monomer density $\phi(r)$. The work extends straightforwardly to cases where $F_{ex}$ depends on more than one weighted density of $\rho(N,r)$, for example the case of `chemical polydispersity' (a model of random copolymers) in which chains have the same length but varying interaction parameter $\chi$ \cite{pago2}. (This requires two weighted densities \cite{long}.) Polydisperse polymer mixtures of several chemical types can likewise be included; the solvent can be made polymeric or removed if desired (by imposing $\phi=1$ as a constraint).

The work of Ref.\cite{pagona} shows that excess free energies of up to four moments, at least, can be treated comfortably, which is probably more than required for most interfacial problems involving dense polymers.

\section{Conclusion}

We have introduced, within a Flory-Huggins framework, a model density functional for polydisperse polymers. This should be valid quantitatively in weak segregation and qualitatively for moderate segregation. By representing chains as a system of spherically symmetric, nondeformable but interpenetrating particles, the functional is able to reproduce the known limiting forms for RPA and square gradient theories in the monodisperse limit. To study the interface thermodynamics of coexisting phases, the closure that we obtain at first-moment level (eq.\ref{eqn:autonomous}) ensures numerical tractability for our model functional, eq.\ref{eqn:polyfunctional}. This closure is inherited by the square gradient expansion of our functional, which has the standard polydisperse form (eq.\ref{eqn:fullsquare}); but in this context the existence of a closed differential equation for the monomer density has perhaps been overlooked. The practical utility of eq.\ref{eqn:polyfunctional} was demonstrated by applying it to find interfacial tensions and density profiles for a polydisperse dense polymer solution at coexistence in a poor solvent.

\acknowledgments
We thank C. Rasc\'on, N. Clarke, T. McLeish, P. Olmsted and P.B.  Warren for illuminating discussions. Work funded in part under EPSRC GR/M29696.



\begin{thebibliography}{0}
\bibitem{warren} WARREN P.B, {\it Phys. Rev. Lett.}, {\bf 80} (1998) 1369.
\bibitem{sollich} SOLLICH P. and CATES M.E., {\it Phys. Rev. Lett.}, {\bf 80} (1998) 1365.
\bibitem{long} SOLLICH P., WARREN P.B and CATES M.E., {\it Adv. Chem. Phys.},
{\bf 116} (2001) 265.
\bibitem{pagona} PAGONABARRAGA I., CATES M.E. and ACKLAND G.J., {\it Phys. Rev. Lett.}, {\bf 84} 911 (2000).
\bibitem{footfactor}
The structure factor $S_0(q)$ for the ideal gas of Gaussian chains is 
the Fourier transform of $\langle\phi(r)\phi(0)\rangle-\phi_0^2$. Since 
chains are uncorrelated, the latter is $\phi_0$ times the monomeric 
pair distribution function within a single chain, whose Fourier 
transform is $Ng_D$.
\bibitem{RPA} LEIBLER L., {\it Macromolecules}, {\bf 13} (1980) 1602.
\bibitem{jodeg} DE GENNES P.-G, {\it J. Phys. (Paris) Lett.},{\bf 38L} (1977) 441; JOANNY J.F. and LEIBLER L., {\it J. Phys. (Paris)}, {\bf 39} 951 (1978).
\bibitem{critique} ZENG X.C., OXTOBY D.W., TANG H. and FREED K.F., {\it J. Chem. Phys.},{\bf 96} (1992); MCMULLEN W.E. and TRACHE M., {\it J. Chem. Phys.},{\bf 102} (1995) 1449 (1995).
\bibitem{broseta} BROSETA D., FREDRICKSON G.H., HELFAND E. and LEIBLER L., {\it Macromolecules},{\bf 23} (1990) 132.
\bibitem{warrenpccp} WARREN P.B, {\it Phys. Chem. Chem. Phys.}, {\bf 1} (199) 2197.
\bibitem{hansen} LOUIS A.A., BOLHUIS P.G., HANSEN J.P. and MEIJER E.J., {\it Phys. Rev. Lett.},{\bf 85} (2000)  2522.
\bibitem{footform} The form factor (kernel) $w$ is the monomer-to-centre 
correlator whose total integral in real space is $N$ 
. For our nondeformable, 
deterministic objects the monomer-monomer correlator $s_0$ follows from $w$ 
by integrating over the centre's position, giving $s_0(q) = w(q)^2/N$. The 
factor $N$ here arises because the monomer-centre and centre-monomer 
correlator integrate to unity and to $N$ respectively, but are (by 
Bayes theorem) identical up to this multiplicative constant.
\bibitem{doi} DOI M. and EDWARDS S.F., {\em The Theory of Polymer Dynamics}, (Clarendon Pr., Oxford) 1986.
\bibitem{pago2} PAGONABARRAGA I. and CATES M.E., {\it in preparation}.
\bibitem{huang} HUANG C. and OLVERA DE LA CRUZ M., {\it Macromolecules}, {\bf 27} (1994) 4231.
\bibitem{clarke} CLARKE N., {\it Eur. Phys. J. E}, {\bf 4} (2001) 327; SCHICHTEL T.E. and BINDER K., {\it Macromolecules}, {\bf 20} (1987) 1671.
\end{thebibliography}
\end{document}